\documentclass{vgtc}                          

\ifpdf
  \pdfoutput=1\relax                   
  \usepackage{graphicx}                
  \DeclareGraphicsExtensions{.pdf,.png,.jpg,.jpeg} 
\else
  \usepackage{graphicx}                
  \DeclareGraphicsExtensions{.eps}     
\fi%

\graphicspath{{figures/}{pictures/}{images/}{./}} 

\usepackage{microtype}                 
\PassOptionsToPackage{warn}{textcomp}  
\usepackage{textcomp}                  
\usepackage{mathptmx}                  
\usepackage{times}                     
\usepackage{cite}                      
\usepackage{tabu}                      
\usepackage{booktabs}                  

\onlineid{0}

\vgtccategory{Research}

\vgtcinsertpkg

\title{Visualizing Comparisons of Bills of Materials}

\author{Rebecca Jones\thanks{e-mail: rebecca.d.jones@pnnl.gov} %
\and Lucas Tate\thanks{e-mail: lucas.tate@pnnl.gov}}
\affiliation{\scriptsize Pacific Northwest National Laboratory}

\abstract{The complexity of distributed manufacturing and software development coupled with the increasing prevalence of cyber and supply chain attacks necessitates a greater understanding of the hardware and software components that comprise equipment in critical infrastructure. When a vulnerability in a single software library can have disastrous consequences, being able to identify where that library may exist in equipment or software becomes a prerequisite for protecting the overall infrastructure. This need has sparked a large effort around the development and incorporation of bill-of-materials (BOM) into security, asset management, and procurement practices to aid in mitigating, and responding to future attacks.
While much of the current research is devoted to creating BOMs, it is equally important to develop methods for comparing them to answer questions, such as: How has my software changed? Are two pieces of equipment equivalent? Does this piece of equipment that just arrived match my historical information? In this work, we demonstrate how BOMs can be represented by graph structures. 
We then describe how these structures can be fed into a graph comparison algorithm to produce a novel interactive visualization that allows us to not only identify differences in BOMs but show exactly where they are in the product.
} 

\CCScatlist{
  \CCScatTwelve{Security and Privacy}{Formal Methods and Theory of Security}{Security Requirements}{};
  \CCScatTwelve{Human-centered computing}{Visu\-al\-iza\-tion}{Visualization Techniques}{Graph Drawings}
}

\begin{document}



\maketitle


\section{Introduction}\label{sec:intro}
Protecting critical infrastructure from cyber attacks, natural disasters, and other disruptions is a priority of the U.S. Government.
Critical infrastructure includes providing electricity to homes and businesses, supplying natural gas for heating, and producing renewable energy sources.
A loss of these services, as seen in the Solarwinds supply chain attack in 2020 \cite{temple-raston_worst_2021}, Texas snowstorm of 2021 \cite{national_weather_service_valentines_nodate}, and the Colonial Pipeline cyber incident of 2021 \cite{office_of_cybersecurity_energy_security_and_emergency_response_colonial_nodate}.
In May 2021, the President of the United States issued an executive order to improve the country's cyber security \cite{house_executive_2021}. 
As part of that order, every piece of software sold to the U.S. government must be accompanied by a software bill of materials (SBOM). 
A BOM is a detailed list of the components in the system and can describe hardware, software, operations, and Software as a Service (SAAS).
The information in the BOM can be used to identify obsolete software as well as highlight potential susceptibility to publicly reported vulnerabilities  \cite{SBOM}. 
Due to the mandate, industry has been exploring the generation of BOMs for their products.

The construction of BOMs today remains an inexact science for numerous reasons \cite{wright_ginger_s4_2023}. 
Some of that variation results from a lack of standardization. 
A primary reason for this is that there are currently competing formats and standards. 
BOMs also vary greatly depending on whether they were produced by a first-party such as the author/manufacturer with complete knowledge or by a third-party with incomplete knowledge. 
A current lack of mature tooling also increases the difficulty of reliably reproducing BOMs, particularly when looking at hardware BOMs which are often constructed manually.
Recorded names or strings can vary widely due to convention,  transcription, or spelling errors.
Other differences can arise based on varying levels of completeness or depth (was every integrated circuit and stop accounted for, or every resistor soldered to the board recorded?). 
Beyond hardware or software components, the relationships linking them together can also be defined in a variety of ways. 
Relationships can be be implied by a nesting structure, described explicitly, represented by a diagram, or possibly even omitted altogether.

Variation can also describe actual differences in composition, and that is exactly what BOMs are designed to capture. 
These differences could be alternative components that were used because they were cheaper, or even a component that had to be replaced because it's been operational for $15$ years. 
Other differences might describe variations across a family of products or even the presence of counterfeiting.
While comparing the competing standards is out of the scope of this paper, the inherent variability in BOMs necessitates tools that allow us to perform comparisons.
The focus of our research is to provide an interactive visual comparison that effectively communicates how two BOMs may be similar or dissimilar to provide valuable insight and help to narrow subsequent analysis.

Current BOM comparison methods include using Excel or proprietary software such as
Oracle Apps\footnote{\url{https://docs.oracle.com/cd/A60725_05/html/comnls/us/bom/bomtas12.htm}}, 
Unisoft\footnote{\url{https://www.unisoft-cim.com/bom-comparison-method-1.html}},
ERPNext\footnote{\href{https://docs.erpnext.com/docs/v12/user/manual/en/manufacturing/bom-comparison-tool}{ERPNext}}.
These tools are limited in the types of BOMs they accept and the data displayed, which does not necessarily include visualizations.
Often, set comparisons are used, which lose the information of how the hardware or software is connected. 
They also focus on evaluating the differences between BOM versions and not necessarily distinctly different BOMs.

To account for the relationships between objects, we convert a BOM into a graph, which we can then easily compare and visualize.
Traditional graph comparisons focus primarily on the structure of the graph but fail to take advantage of other information available within a BOM. 
To compare BOMs accurately, we need a method that allows us to incorporate important component information such as names, hashes, or versions, as well as structural information describing how those components fit together.
We create a mapping that describes how the objects/components in one graph map to the components/objects in the other graph based on a depth-first search algorithm. 
When constructing the mapping, we can choose which information we want to consider (e.g. name, hash, name and hash) as well as whether the mapping should utilize exact or fuzzy matches. 
Fuzzy matching can be useful in instances where names or strings might have spelling or transcription errors, and can suggest where nodes in the graphs might have intended to reference the same component.
Once constructred, the mapping is then used to combine the BOM graphs into a single merged graph.

\subsection{Outline}
In this paper, we will demonstrate our method for comparing BOMs and show how visualizing this approach can help analyze BOM differences.
We will start out by discussing BOMs and how they can be  represented as graphs, as well as current research in \autoref{sec:background}.
\autoref{sec:merging} contains the method for combining BOMs, and \autoref{sec:example} demonstrates the visualizations in two different examples.
Lastly, we'll end with our conclusion in \autoref{conclusions}.

\section{Background}\label{sec:background}
A bill of materials (BOM) is a list of all parts needed to produce a product.
For each part in the finished product, the BOM can store various attributes like version, manufacturer, vendor, hash, package URL (PURL), and location. 
At this time, there is no single industry standard for generating BOMs. 
SBOMs have received more attention and presently, the three most widely used open source standards are CycloneDX \cite{owasp}, Software package data exchange (SPDX) \cite{SPDX}, and Software identification (SWID) tags \cite{computer_security_division_software_2018}.
In addition to including information on the components, some BOMs also contain information on how the components are connected.
This can be a separate document like a drawing or a list of relationships.
The relationships in a BOM are important for identifying vulnerabilities, since connections may exacerbate or nullify a vulnerability.

\subsection{Bill of Materials as Graphs}\label{ssec:bom_as_graph}
Many of the BOM comparison tools listed in \autoref{sec:intro} use set comparisons on the BOM objects.
This results in a loss of information about how the components in the BOM are connected and is why relationships are needed in a BOM.
For instance, a 12-pin port appearing in two hardware BOMs could be connected to different circuit boards in the actual products.
There is also no guarantee that objects in BOMs have the same names, unique ids, or other metadata.
This emphasis on needing connections naturally leads to representing BOMs as graphs.

\begin{figure}[hp]
    \includegraphics[width=.5\textwidth]{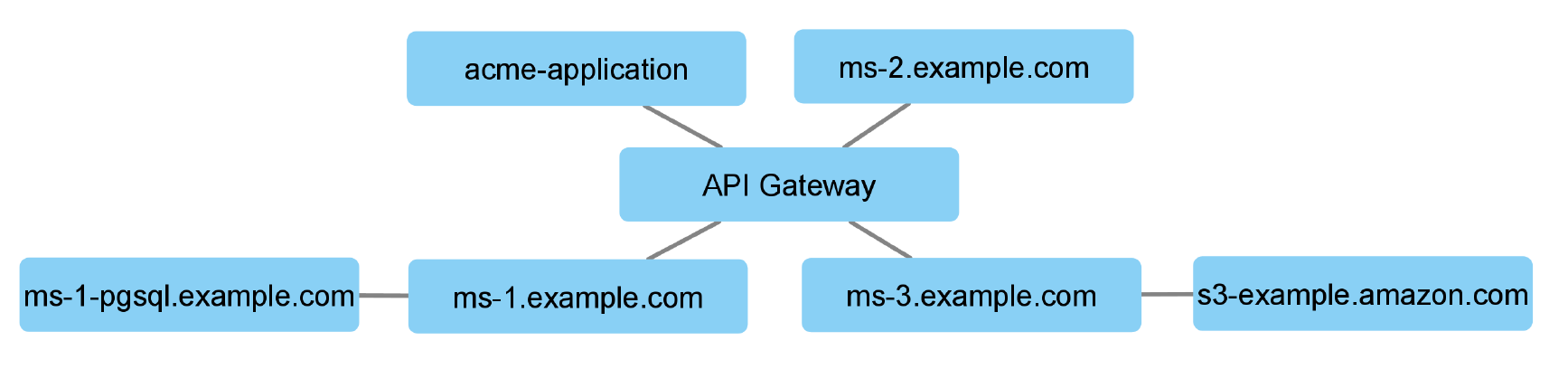}%
    \caption{Visualization of an open-source Saas BOM. Created using Cytoscape \cite{shannon_cytoscape_2003}.}
    \label{fig:sbom}
\end{figure}

A graph is a mathematical object that represents how things are connected.
Each component of the BOM becomes a node in the graph while the relationships determine the edges. 
Edges are created between components when a component possesses some relationship (e.g. contains, uses, imports, includes, etc.) to another component.
For example, if a software file imports another software file, then there would be an edge from one file node to the other file node. 
A hardware example is when a chip is soldered onto a circuit board, so there is an edge from the circuit board node to the chip node in the graph.
\autoref{fig:sbom} shows the graph for an open-source example \href{https://github.com/CycloneDX/bom-examples/tree/master/SaaSBOM/apigateway-microservices-datastores}{SAAS BOM} stored in a json file \cite{cycloneDx_examples}.
Each of the nodes in the graph represent one of the seven services listed in the BOM, uniquely identified by the ``bom-ref".
The edges are created from the dependencies key.
Each entry in contains a ``ref" and a ``depends on" key which list the ``bom-ref" ids that are connected.

Notice that since there could be duplicated parts in a system, there could be similar components in a BOM, e.g. multiple chips of the same type.
In a BOM, each component has a unique ID, and different relationships to distinguish between the different components.
Also, the metadata included in the BOM could be added to the graph as node attributes, features that are attached to a node. 
For instance, each node could have a vender attribute that lists the product vendor.

\subsection{Previous Research}

As mentioned, a graph is a natural choice to represent a BOM that contains relationships \cite{wang_new_2013, guoli_analysis_2003, schmidt_michael_graph-based_2017, carmody_building_2021}.
Hypergraphs can also be used to represent certain BOMs  \cite{nagy_hypergraph-based_2022}, although we note that not all BOMs have the structure needed to created a hypergraph.
Graph databases are relational databases that store node data and relationships, i.e., graphs, instead of tables and have been suggested for storing BOM data \cite{aydin_effective_2005}.
A high profile commercial graph database, Neo4j, is used by the Army to manage their BOMs \cite{army_neo4j} quickly and efficiently \cite{guia_graph_2017}.
In comparing BOMs, previous research has attempted to answer the question, ``How similar are two BOMs?"
This has often been answered with a distance metric \cite{shih_product_2011, kashkoush_product_2016, romanowski_comparing_2005}.
There is some work that looks at combining multiple BOMs into graphs:  aggregating multiple BOMs into one network to calculate network properties on the combined graph \cite{cinelli_network_2017}, and  demonstrating graph matching methods to match pairs of BOMs to reduce production time \cite{kashkoush_matching_2013}.

There is more work on comparing general graphs.
Methods for this include calculating distance metrics which determine how far apart graphs are in some metric space \cite{amenta_case_2002}, clustering which looks at structural differences\cite{archambault_structural_2009}, deep learning \cite{han_visual_2020, fujiwara_visual_2020}, and node correspondence algorithms \cite{tantardini_comparing_2019}. 
We use the latter, in which nodes from one graph are mapped to another graph.
This can be done where the node mapping is already known such as a cut distance \cite{liu_cut_2018} or where the node mapping is not known.
When the node mapping is unknown, it is referred to as an unknown node correspondence, or UNC.
One UNC approach is called alignment, where the mapping between graphs is created by optimizing a similarity function over the graph \cite{kuchaiev_topological_2010}.
Many of the unknown node correspondence algorithms only use the graph structure and some are restricted to trees \cite{ munzner_treejuxtaposer_2003, priel_vectorial_2022}.
The node mapping can also include the node features (in our case BOM metadata) in the similarity score calculation \cite{euzenat_similarity-based_2004}.

Visualizing graphs, especially large graphs, is challenging.
The superposition approach overlays the two graphs \cite{gleicher_visual_2011}.
This is similar to our method, in which we merge the two graphs into one and display the combined graph.
Because there is more node metadata than can reasonably be presented at one time, a useful visualization for analysis will be interactive and allow users to manually compare the graphs \cite{huerta-cepas_ete_2010, alper_weighted_2013, hascoet_interactive_2012}.
Most BOM visualization techniques leverage tables to display the BOM differences \cite{freire_manynets_2010, shilovitsky_openbom_2022}.
Often, these are compared using set comparisons, which leave out information about relationships, but they can facilitate multiple comparisons at once \cite{alsallakh_state---art_2016}.
Visualizing sets can incorporate structure, but it's not as easy to see the differences \cite{telea_code_2008}.
SVG is a visualization tool that uses the Hungarian Assignment algorithm to map nodes and can include semantics like node attributes \cite{andrews_visual_2009}.
It requires node similarities to be calculated beforehand while our focus is calculating the mapping.

\section{Combining Two BOM Graphs}\label{sec:merging}
Our method of combining BOMs uses an unknown node mapping, where each node in one graph is mapped to a unique node in the second graph, if possible.
To do this, we implement a depth-first search algorithm \cite{tarjan_depth-first_1972}.
Depth first search is a traversal algorithm, meaning it visits each node in the graph exactly once and runs in linear time with the number of nodes in the graph.
At each node, a similarity matching is done against the nodes in the other graph.
The matching can be done on any number of node attributes and can be any string similarity metric, such as exact matching or Jaro-Winkler \cite{cohen}.

\begin{figure}[hp]
    \includegraphics[width=.45\textwidth]{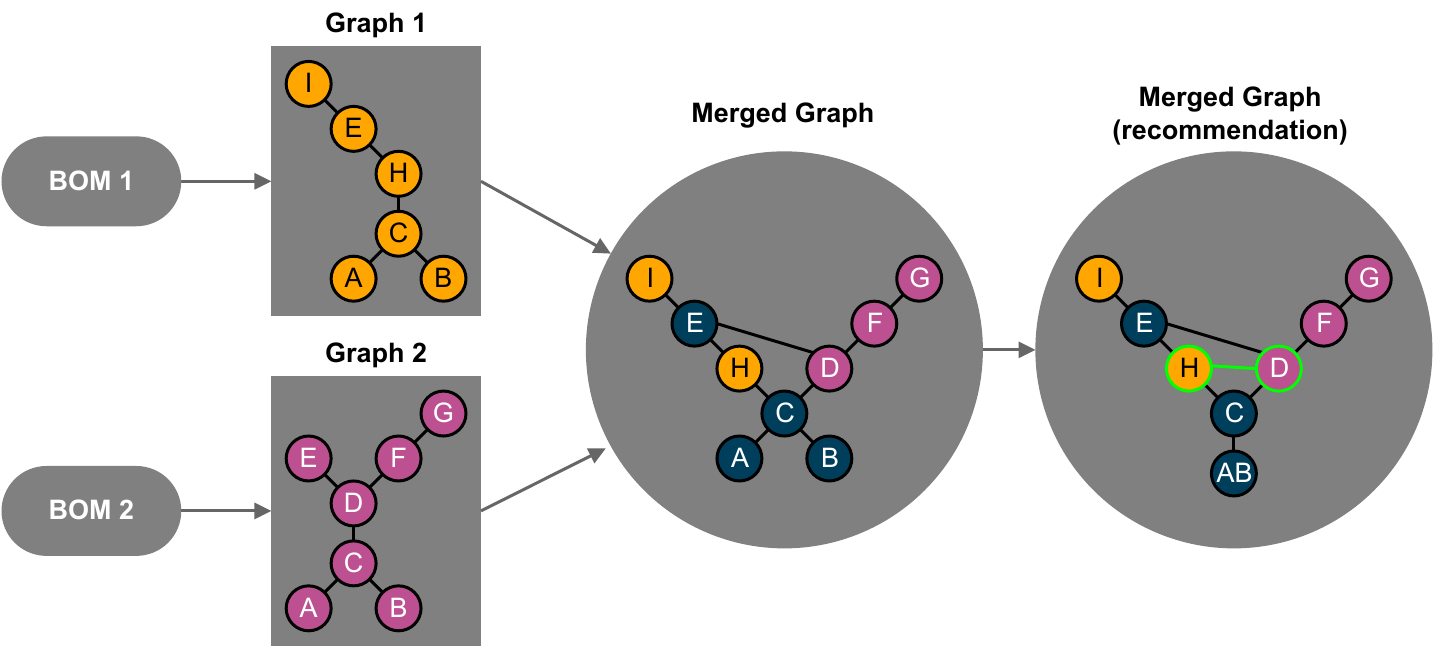}%
    \caption{The process of combining BOMs into a single graph. Blue nodes indicate a successful mapping between the graphs. Pink nodes indicate that the node is only in Graph 1. Yellow nodes mean that the nodes appear in Graph 2. Green edges indicate that the connected nodes might be the same; they have similar neighbors and have node attributes that meet some similarity threshold.}
    \label{fig:flow}
\end{figure}

The process is outlined in \autoref{fig:flow}.
First, two BOMs are turned into graphs, as illustrated in \autoref{fig:sbom}.
A starting node in Graph $1$ is chosen that has a known match in Graph $2$, such as $C$.
Graph $1$ is traversed, mapping nodes in Graph $1$ to nodes in Graph $2$ that have similar edges and attributes (or names as in the example).
Then the graphs are combined, using the node mapping.
This yields the first merged graph in \autoref{fig:flow}.
Throughout the paper, blue represents nodes that have a mapping from Graph 1 to Graph 2, yellow represents nodes in Graph 1, and pink represents nodes only in Graph 2.

For further analysis, we include an option to predict nodes that might be the same based on fuzzy string matching and graph neighbors, as seen in the far right circle of \autoref{fig:flow}.
This is done by rerunning the depth first search algorithm with any fuzzy string matching algorithm on the nodes that were not mapped, $D$, $F$, and $G$ in our example. 
Assuming $H$ and $D$ are close through some string similarity, since they have the same neighbors, a green edge is shown between them, depicting a possible match.
For graphs where large number of nodes result in severe overplotting, we include an option to condense the merged graph by combining leaf nodes that appear in both graphs into a single supernode, as with nodes $A$ and $B$.

An interactive .html version of the combined graph is saved, as well as a .gml, so the merged graph can be imported into most graph visualization tools.
Our method is an end-to-end system, taking in graphs generated from BOMs, merging the graphs, and creating a visualization.
Some of the benefits of our method are that we allow any type of graph (we're not restricted to trees), that we include the node attributes and structure in the node mapping, we allow for fuzzy string matching, and we include interactive visualizations.

\section{Use Cases}\label{sec:example}
\begin{figure}[h]
\begin{minipage}[b]{0.45\linewidth}
\centering
\includegraphics[width=\textwidth]{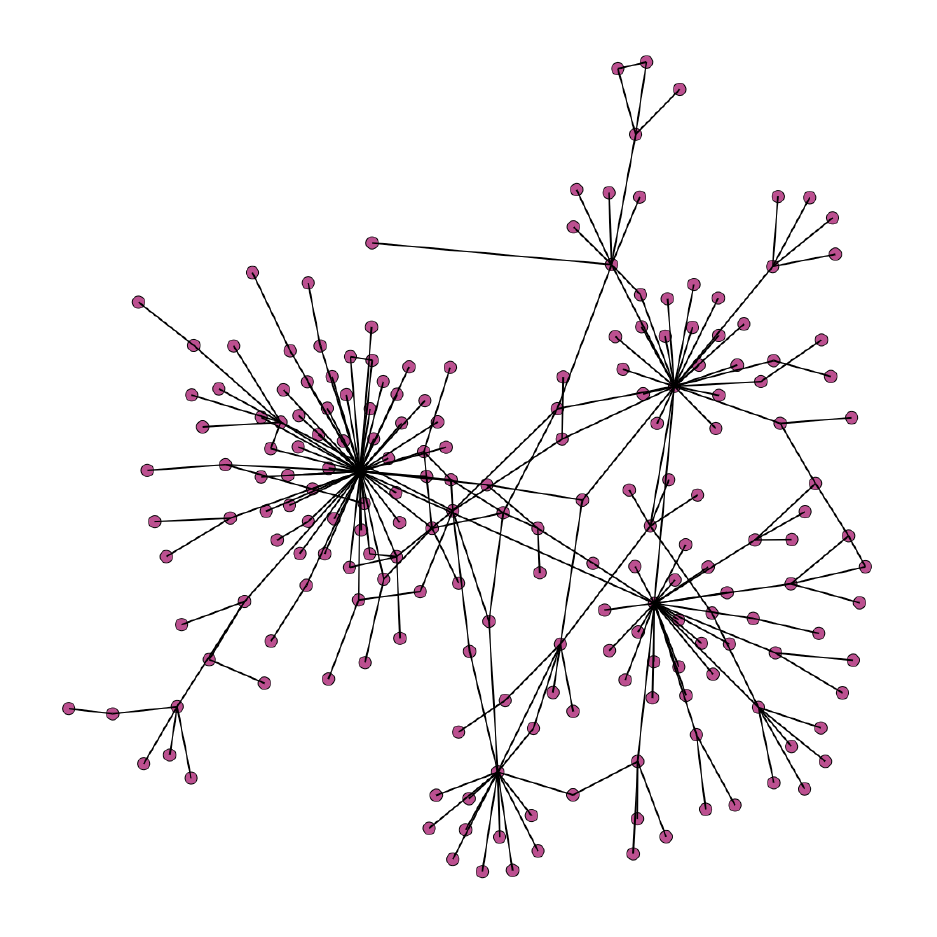}
    \caption{Visualization of proton-bridge-v1.6.3 BOM.}
    \label{fig:v6}
\end{minipage}
\hspace{0.5cm}
\begin{minipage}[b]{0.45\linewidth}
\centering
\includegraphics[width=\textwidth]{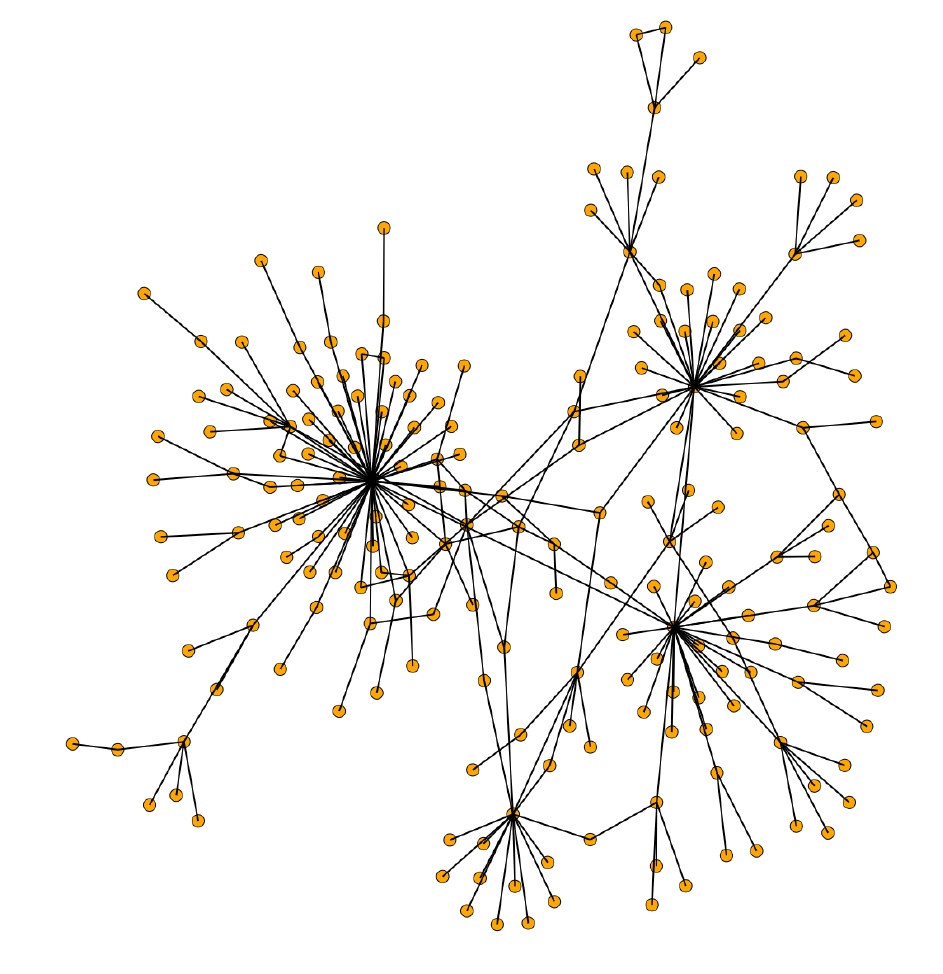}
    \caption{Visualization of proton-bridge-v1.8.0 BOM.}
    \label{fig:v8}
\end{minipage}
\end{figure}

\begin{figure}[h]
\centering
    \includegraphics[width=.4\textwidth]{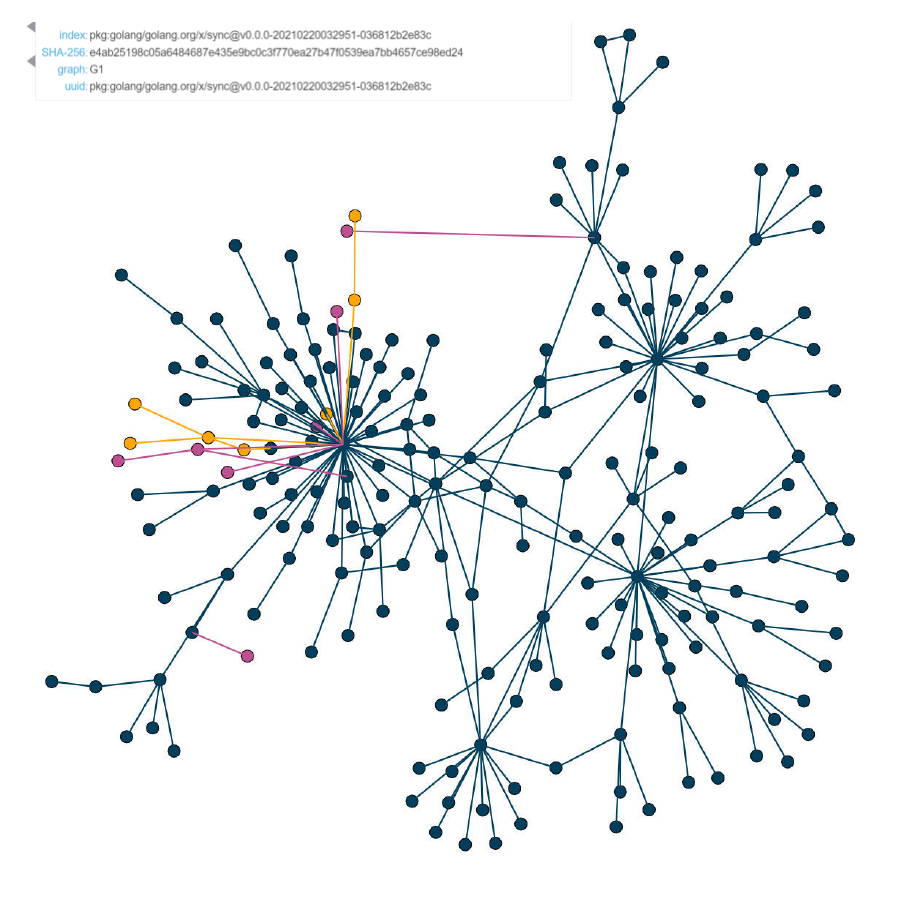}%
    \caption{Visualization of the combined proton-bridge graph. Blue nodes represent components with the same SHA-256 hash and edges. Pink nodes appear only in version 1.6.3, and yellow nodes only appear in version 1.8.0. The green edge is reflects relationships to the highlighted node. Node information for the selected node is displayed in the top left corner.}
    \label{fig:proton_comb}
\end{figure}

To demonstrate the usefulness of this method, two illustrative use cases are provided.
The first use case looks at two versions ($1.6.3$ and $1.8.0$) of an open-source example \href{https://github.com/CycloneDX/bom-examples/tree/master/SBOM/proton-bridge}{SBOM} for proton-bridge.
The graphs, displayed in \autoref{fig:v6} and \autoref{fig:v8}, were created in the same manner as the example in \autoref{ssec:bom_as_graph} where components become nodes in the graph and the relationships are created from the dependencies.
The node mapping was done with an exact match on the SHA-256 hash.
So nodes are considered equal if they have the same hash and the same neighbors.
By looking at the merged graph shown in \autoref{fig:proton_comb}, we can easily see the differences between the two versions through the pink (version $1.6.3$) and the yellow (version $1.8.0$) nodes. 
When a user hovers over a node, its edges turn green to highlight the node's neighbors, and the node attributes are listed at the top.
This way, a user can manually select and compare node information to identify discrepancies and possible differences.


From a cyber perspective, only the yellow nodes, nodes in the newer software version, will drive new behaviors or introduce new vulnerabilities, which should allow us to narrow our focus when analyzing new BOMs. 
The utility of comparing within a version controlled repository is diminished since much of the insights can be reproduced using version control software, so it's important to note that this approach is going to be agnostic to varied node definitions (e.g. files, directories, functions, containers, etc.) or relationship definitions (e.g. contains, imports, installs, downloads, etc.) and furthermore allows the user to choose which node attribute(s) should be utilized for the mapping (e.g. names, hashes, etc.).

\begin{figure}[h]
\begin{minipage}[b]{0.45\linewidth}
\centering
\includegraphics[width=\textwidth]{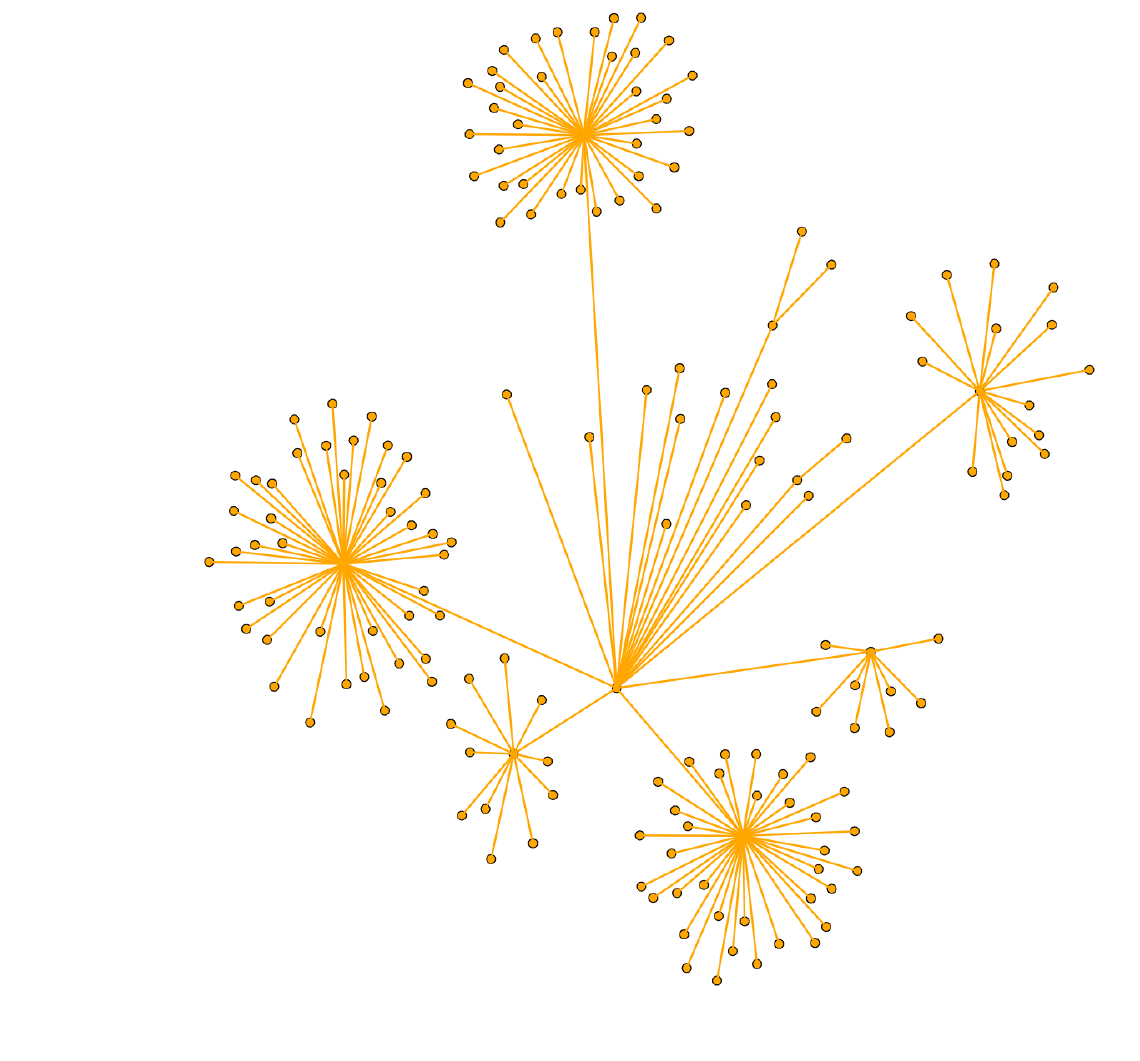}
\caption{First HBOM Graph}
\label{fig:G1}
\end{minipage}
\hspace{0.5cm}
\begin{minipage}[b]{0.45\linewidth}
\centering
\includegraphics[width=\textwidth]{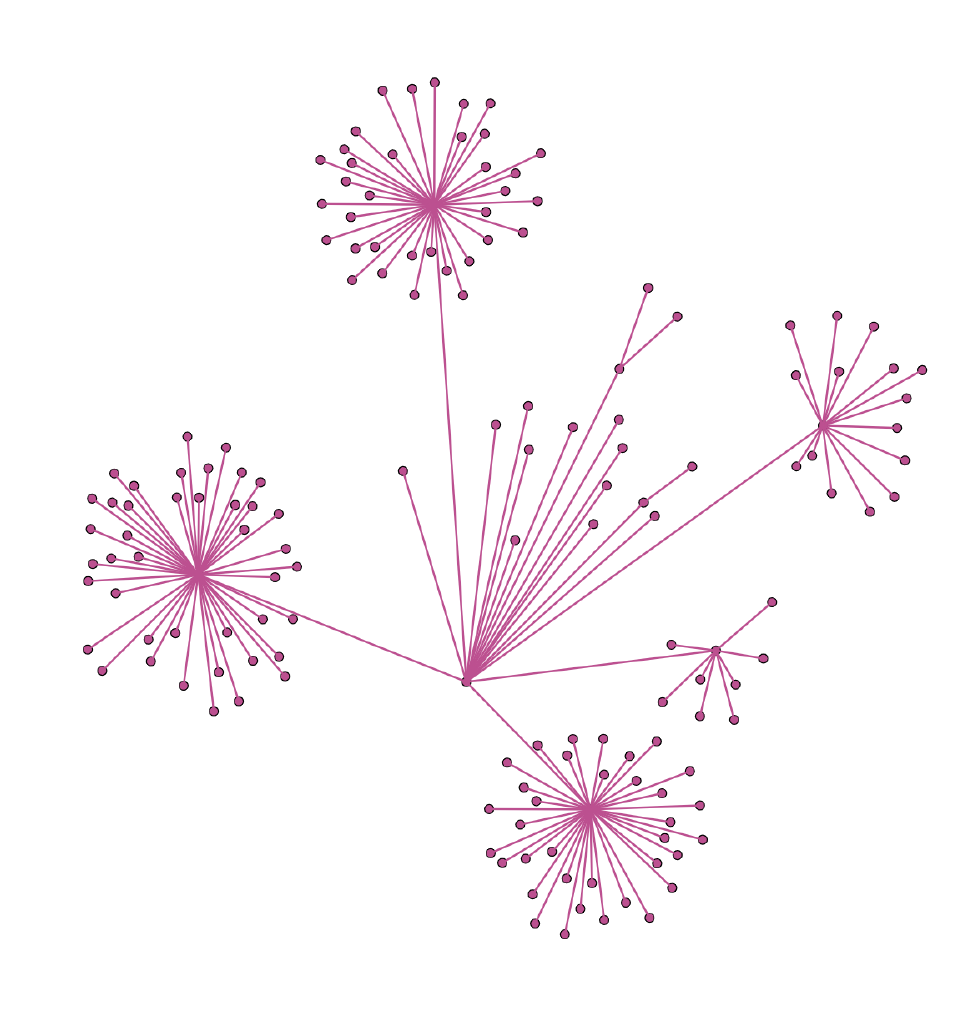}
\caption{Second HBOM Graph}
\label{fig:G2}
\end{minipage}
\end{figure}

\begin{figure}[h]
\centering
\includegraphics[width=0.35\textwidth]{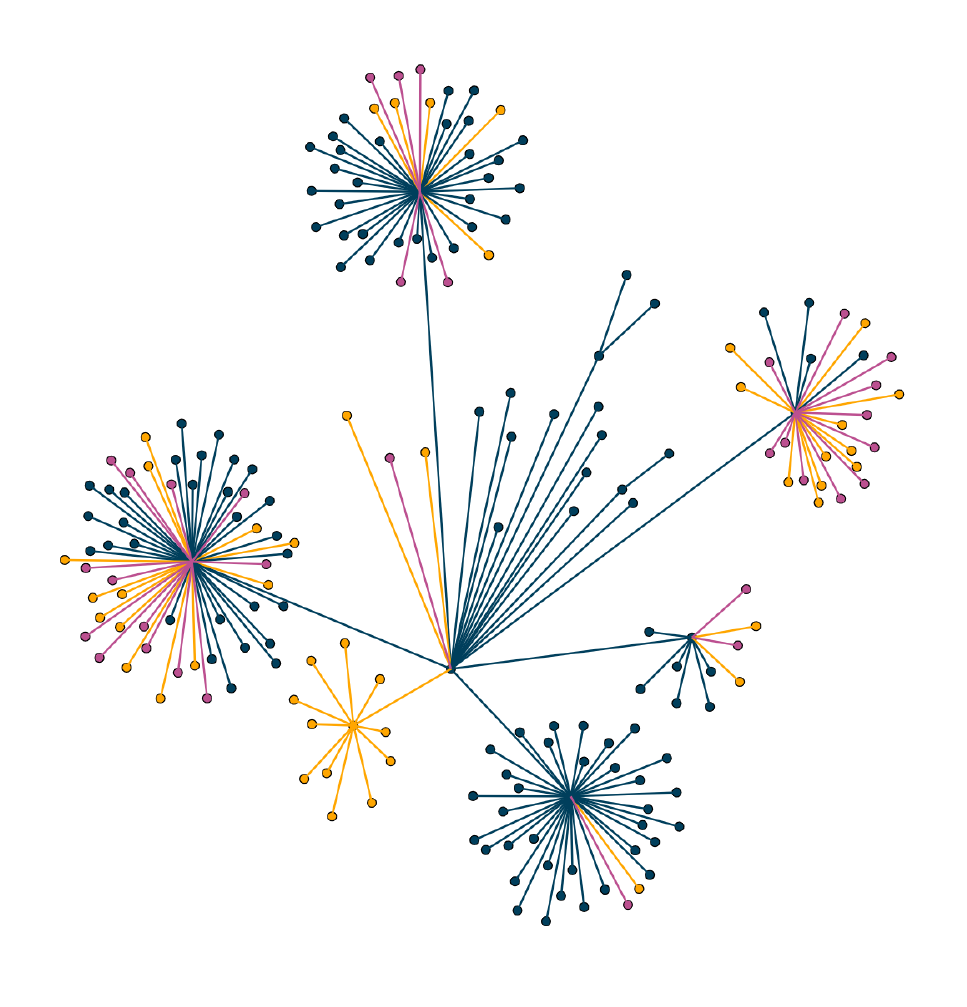}
\caption{Comparison of the exact component names between HBOMs generated on two distinct hardware devices of the same model. Blue nodes indicate mapped nodes in both graphs.}
\label{fig:combined}
\end{figure}

\begin{figure}[h]
\centering
\includegraphics[width=0.35\textwidth]{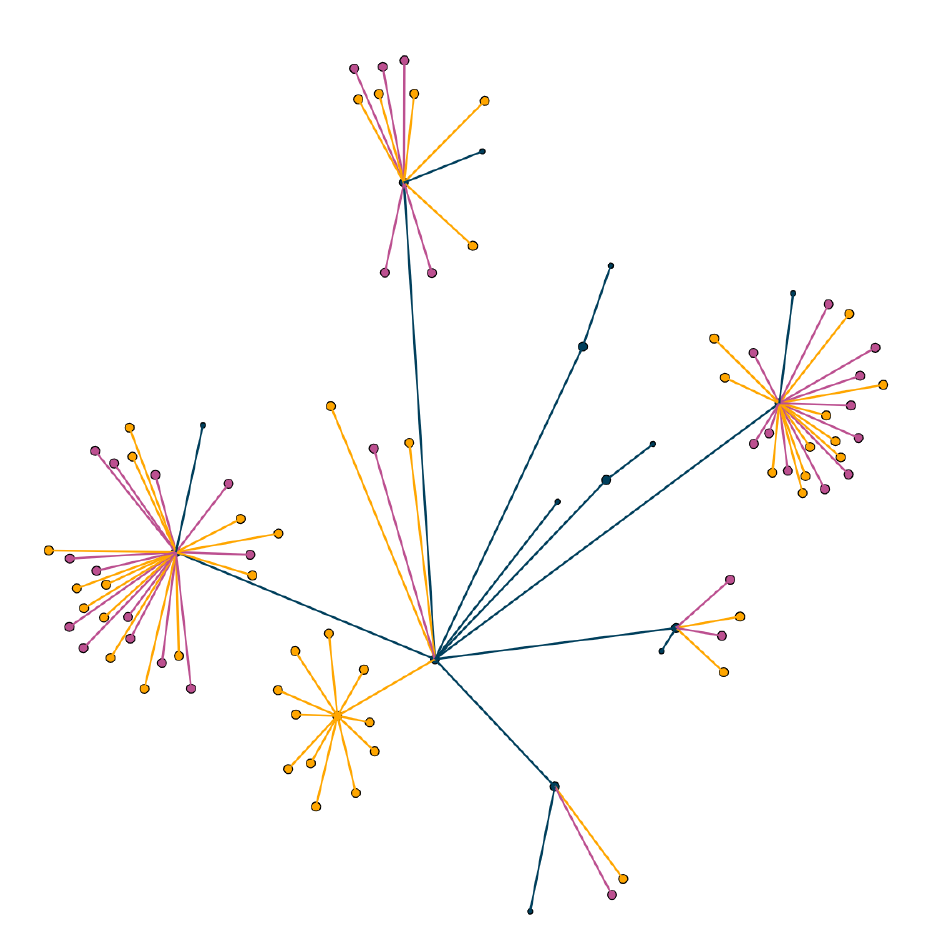}
\caption{Combined HBOM graph with leaf nodes collapsed.}
\label{fig:collapsed}
\end{figure}

\begin{figure}[h]
\centering
    \includegraphics[width=0.35\textwidth]{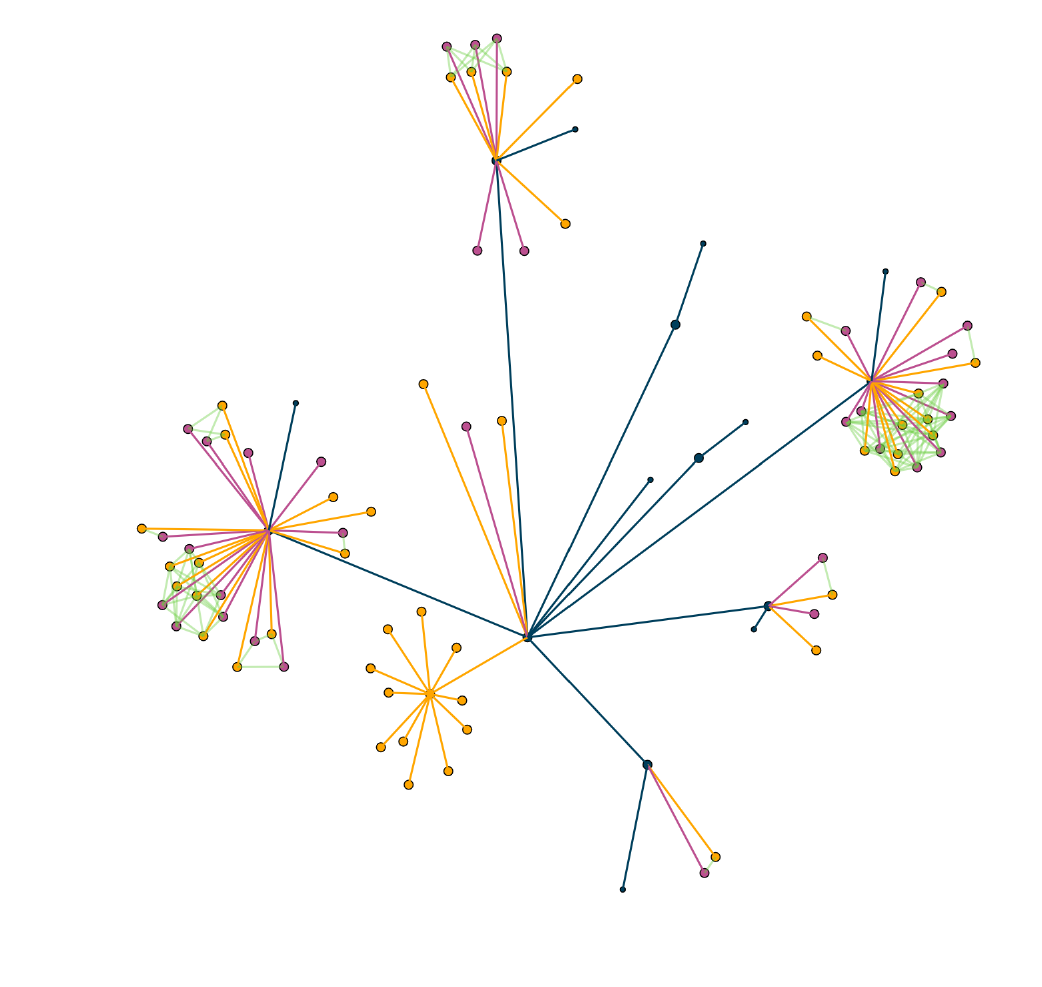}%
    \caption{Combined HBOM graph with suggested mappings shown in green. }
    \label{fig:comparison_7d0_446}
\end{figure}

Next, we look at an actual comparison of two hardware bill of materials (HBOMs) \autoref{fig:comparison_7d0_446}. 
In this real example, the HBOMs were furnished by the same entity and generated for two devices  of the same model, represented by \autoref{fig:G1} and \autoref{fig:G2}. 
Nodes represent hardware components in the device and relationships show how the components are physically connected.
A first pass of the algorithm used exact matching on the node names to generate the initial node mapping, creating the merged graph shown in \autoref{fig:combined}.
Blue nodes once again represent components in both BOMs.
This means that the components have the exact same name and neighbors in both devices.
We condensed the merged graph by collapsing blue leaf nodes into supernodes, as seen in \autoref{fig:collapsed}.
In a small graph like this one, collapsing the nodes is not needed, but when the graph becomes too large to easily display, collapsing leaf nodes can simplify the graph and allow the differences to be easily seen.
A second pass of the algorithm used a fuzzy match on the names to draw green edges, see \autoref{fig:comparison_7d0_446}, which suggests where components may be similar.

At first glance we can see that there is a central blue structure in the graph which suggests the core structures of the two devices are the same. 
This isn't especially surprising since we are comparing two devices of the same model. 
From there we can visually make some additional observations. 
First, there is a yellow structure jutting from the center to the lower left. 
This suggests (and was confirmed) to be an additional circuit board in one of the devices accompanied by the corresponding mounted components.



Second, the green edges point to instances of similarly named components at the same structural location in each device. 
By manually inspecting the nodes connected by green edges, subject matter experts (SMEs) were able to view the metadata and compare the two nodes.
This  quickly revealed  similarly named components, spelling errors, and subtle differences in the convention used to record names.
For example, $AS298$ mapped to $A5298$, likely a transcription error switching the $S$ to a $5$.
The denser green webs form where there are multiples of a component on one device that got mapped to multiples of a component on the other device and the names are very similar.

The interactive visualization showed marked benefits over the baseline approach of set comparisons and greatly increased the ability of SMEs to explore, explain, and  respond appropriately to discrepancies that were identified.
Along with discovering the additional circuit board, SMEs were able to identify some major inconsistencies, likely spelling or transcription errors as described above, that had previously been undetected using this tool.
With subsequent use of the visualization SMEs were quickly able to form an intuition about the differences in similar graphs, greatly increasing the speed and efficiency of the BOM comparisons.



\section{Conclusion}\label{conclusions}
The increasing prevalence of BOMs requires new tools and methods to leverage them.  
While policy is largely driving the creation and adoption of BOMs, methods for analyzing them have lagged.
Our method describes a flexible approach to easily explore the similarity or dissimilarity of two BOMs based on any number of metadata attributes. 
By presenting the results in an interactive graph visualization, users are able to quickly identify and explore differences for asset management or security use cases.

\acknowledgments{
The authors wish to thank the Department of Energy (DOE) Cybersecurity, Energy Security, and Emergency Response (CESER) and the Cyber Testing and Resilience of Industrial Control Systems (CyTRICS) Program including Idaho National Laboratory (INL), Lawrence Livermore National Laboratory (LLNL), National Renewable Energy Laboratory (NREL), Oakridge National Laboratory (ORNL), and Sandia National Laboratory (SNL).
}
\clearpage
\bibliographystyle{abbrv-doi}

\bibliography{refs}
\end{document}